\documentclass[11pt]{article}
\usepackage{epsfig} 
\newcommand{\sect}[1]{ \section{#1} \setcounter{equation}{0} }

\newcommand{\pslash}{p \! \! \! /} 
 
\newcommand{\partialslash}{\partial \! \! \! /} 
\newcommand{\half}{\mbox{\small{$\frac{1}{2}$}}} 
\newcommand{\MSbar}{\overline{\mbox{MS}}} 
\newcommand{\MSbars}{\overline{\mbox{\footnotesize{MS}}}} 
\newcommand{\NN}{{\cal N}} 
\newcommand{\OO}{{\cal O}} 

\begin{document}
\title{Four loop $\MSbar$ mass anomalous dimension in the Gross-Neveu model}
\author{J.A. Gracey, \\ Theoretical Physics Division, \\ 
Department of Mathematical Sciences, \\ University of Liverpool, \\ P.O. Box 
147, \\ Liverpool, \\ L69 3BX, \\ United Kingdom.} 
\date{} 
\maketitle 
\vspace{5cm} 
\noindent 
{\bf Abstract.} We compute the four loop term of the mass anomalous dimension
in the two dimensional Gross-Neveu model in the $\MSbar$ scheme. The absence of
multiplicative renormalizability which results when using dimensional 
regularization means that the effect of the evanescent operator, which first 
appears at three loops in the $4$-point Green's function, has to be properly 
treated in the construction of the renormalization group function. We repeat
the calculation of the three loop $\MSbar$ $\beta$-function and construct the
$\beta$-function of the evanescent operator coupling which corrects earlier
computations.

\vspace{-17cm}
\hspace{13.5cm}
{\bf LTH 786}

\newpage 

\sect{Introduction.}

The Gross-Neveu model is a two dimensional asymptotically free renormalizable
quantum field theory whose basic interaction is a simple quartic fermion
self-interaction, \cite{1}. It has been widely studied since its introduction 
in \cite{1}, as it has many interesting properties which are readily accessible
given the space-time dimension the model is defined in. For instance, unlike 
the same interaction in four dimensions it is renormalizable in two dimensions 
and the generation of mass dynamically has been observed and studied in the 
large $N$ expansion, \cite{1}. Moreover, one property of interest is that it 
possesses an $S$-matrix whose {\em exact} form is known, \cite{2,3}, whence the
mass gap is known exactly, \cite{4}, in terms of the basic mass scale of the 
theory, $\Lambda_{\MSbars}$. Aside from these features the model itself 
underpins several problems in condensed matter physics. For instance, in the 
replica limit it is equivalent at the critical point to the two dimensional 
random bond Ising model. (See, for example, the review \cite{5}.) Necessary to 
study the fixed point properties for such physical problems is knowledge of the
renormalization group functions in some renormalization scheme, such as 
$\MSbar$. These have been computed to three loops in $\MSbar$ over a period of 
years. In \cite{1} the one loop $\beta$-function was computed demonstrating 
asymptotic freedom. This was extended to two loops in \cite{6}, whilst the 
three loop $\beta$-function appeared more or less simultaneously in \cite{7} 
and \cite{8}. Though the method of computation in both articles was 
significantly different. For instance, given the quartic nature of the sole 
interaction it can be rewritten in terms of an auxiliary field producing a 
trivalent interaction with the introduction of an auxiliary field. This was the
version of the theory used in \cite{8}, as well as at two loops in \cite{6}, 
not only to deduce the $\beta$-function but also to study the effective 
potential of the auxiliary field at three loops. However, renormalization 
effects will generate a quartic interaction. So \cite{8} had in effect to 
handle the intricate problem of renormalizing a version of the theory with two 
independent couplings. The three loop $\MSbar$ $\beta$-function of the original
theory was eventually extracted when the effect of the newly generated 
interaction was properly accounted for in the renormalization group equations.
By contrast, in \cite{7} the purely quartic version of the theory was treated
with a massive fermion. The agreement of both three loop results was a 
reassuring non-trivial check on the final expression. At four loops only the 
wave function renormalization has been computed in \cite{9} and later verified 
in \cite{10}. Although apparently one loop further than the $\beta$-function
of \cite{7,8}, or mass anomalous dimension, \cite{11}, the fermionic nature of 
the interaction means that the anomalous dimension begins at two loops since 
the one loop snail graph is zero in the wave function channel of the $2$-point
function. Thus in effect the four loop wave function is a computation on a par 
with the three loop mass anomalous dimension and $\beta$-function.

Given the range of problems which the Gross-Neveu model underlies, it is the 
purpose of this article to start the programme of completing our knowledge
of the four loop structure by computing the mass anomalous dimension at this
order in the $\MSbar$ scheme. This may appear to be the least important of the
two outstanding quantities. However, as will become apparent from the 
calculation we will describe, the nature of the model is such that there are
several difficult technical issues to be dealt with en route which do not
arise in other four loop calculations in other important theories. Therefore,
in one sense we are testing the viability of computing renormalization 
constants at four loops in an example in the Gross-Neveu model which contains a
moderate number of Feynman diagrams rather than the $1000$ plus graphs which 
will occur in the $4$-point function renormalization {\em and} with a sensible 
investment of time. Moreover, the experience gained in such an exercise will be
invaluable for any future such coupling constant renormalization. It will 
transpire that we will require properties of Feynman integrals in two 
space-time dimensions higher than the one we compute in, in order to determine 
the final renormalization group function. Whilst this may not be the unique way
to determine this, it will suggest the importance of the structure of higher 
dimensional Feynman diagrams in complementing practical lower dimensional 
calculations and potentially {\em equally} vice versa at high loop order. Our 
main tool of computation will be the use of dimensional regularization in 
$d$~$=$~$2$~$-$~$\epsilon$ dimensions where the relevant part of the $2$-point 
function is written in terms of basic massive vacuum bubble graphs. Such a 
calculation could only be completed with the use of computer algebra and 
invaluable in this was the symbolic manipulation language {\sc Form},
\cite{12}. At this level of loop order automatic Feynman diagram computation
using computers is the only viable way of keeping a reliable account of the
algebra within a reasonable time. Several other computer packages were also 
required.

As this is the first four loop calculation which involves {\em four} terms of
the renormalization group functions, we will have to revisit and redo the
earlier calculations using the same approach consistently as that which will
be used here at four loops. For reasons which will become evident later this 
has entailed us carrying out the full renormalization of the $4$-point function
in the theory with a massive fermion, where the mass will act as a natural
infrared regulator. In \cite{7} and \cite{8} these articles centred on
the derivation of the three loop $\beta$-function itself, which unlike most
field theories, is not the same as fully renormalizing the underlying
$n$-point function. For the Gross-Neveu model this was first observed in
\cite{13,14} with explicit three loop calculations for the full $4$-point 
function given for a massless version of the theory discussed in the series of
articles \cite{15,16,17} and examined in \cite{18} for the massive version. 
However, neither computation was in complete agreement as to the final 
structure of the $4$-point function renormalization. Prior to considering any 
four loop mass anomalous dimension this discrepancy needs to be resolved in one 
way or another which is a secondary consideration for this article. Any four 
loop $\beta$-function computation would also require this resolution but
constructing the mass anomalous dimension in a consistent way is an easier 
environment in which to {\em check} any final explanation.

The paper is organised as follows. We review the current understanding of the
renormalization properties of the Gross-Neveu model in section $2$. Given this
we discuss the three loop vacuum bubble integrals needed for the full three
loop renormalization in section three. There we resolve the discrepancy between
\cite{17} and \cite{18} in the $4$-point renormalization in section $4$. 
Section $5$ extends the discussion of the relevant vacuum bubble computations 
to four loops with the four loop mass anomalous finally being derived in 
section $6$. Concluding remarks are given in section $7$. 

\sect{Preliminaries.}

We turn now to the specific properties of the Gross-Neveu model. The bare
two dimensional Lagrangian is, \cite{1},  
\begin{equation} 
L ~=~ i \bar{\psi}_0^{i} \partialslash \psi^{i}_0 ~-~ m_0 \bar{\psi}^{i}_0 
\psi^{i}_0 ~+~ \frac{1}{2} g_0 ( \bar{\psi}^i_0 \psi^i_0 )^2
\label{baregn} 
\end{equation}  
where the subscript ${}_0$ denotes a bare quantity and $g$ is the coupling
constant which is dimensionless in two dimensions. Unlike \cite{7,11} which
used the symmetry group $O(N)$ we take the $SU(N)$ version of the theory so 
that the fermion field $\psi^i$ is a complex Majorana fermion with the former 
property deriving from the fields taking values in the group $SU(N)$ with 
$1$~$\leq$~$i$~$\leq$~$N$. An advantage of the choice of the group $SU(N)$ is
that the Feynman rule of (\ref{baregn}) involves two terms unlike the three of
the $O(N)$ case. At four loops this reduces the number of terms needed to be
substituted when the Feynman rules are implemented in {\sc Form} which speeds
up the calculation. We choose to work with the massive version where $m$ is the
mass. This is primarily because we will need to redo the three loop 
renormalization of the coupling and the presence of a non-zero mass will ensure
that any resulting divergences are purely ultraviolet and not deriving from 
spurious infrared infinities which could arise when external momenta are 
nullified in the basic divergent $4$-point Green's function. In two dimensions 
the theory (\ref{baregn}) is renormalizable to all orders in the coupling 
constant and is asymptotically free. Specifically we note that the $\MSbar$ 
scheme renormalization group functions of the model, as they currently stand 
are, \cite{1,6,7,8,9,11,17}, 
\begin{eqnarray}
\gamma(g) &=& (2N-1) \frac{g^2}{8\pi^2} ~-~ (N-1)(2N-1) 
\frac{g^3}{16\pi^3} \nonumber \\
&& +~ (4N^2-14N+7)(2N-1) \frac{g^4}{128\pi^4} ~+~ O(g^5) \nonumber \\ 
\gamma_m(g) &=& -~ (2N-1) \frac{g}{2\pi} ~+~ (2N-1) \frac{g^2}{8\pi^2} ~+~ 
(4N-3)(2N-1) \frac{g^3}{32\pi^3} ~+~ O(g^4) \nonumber \\ 
\beta(g) &=& (d-2)g ~-~ ( N - 1 ) \frac{g^2}{\pi} ~+~ 
( N - 1 ) \frac{g^3}{2\pi^2} ~+~ ( N - 1 ) ( 2N - 7 ) \frac{g^4}{16\pi^4} 
\nonumber \\
&& +~ O(g^5)
\label{threerge}
\end{eqnarray}  
where $\gamma(g)$, $\gamma_m(g)$ and $\beta(g)$ are respectively the field and
mass anomalous dimensions and the $\beta$-function. Their formal definitions
will be discussed later. Although several terms were determined for the $O(N)$
version of the model, we have converted the previous computations to the
$SU(N)$ model whence the free field case emerges when $N$~$=$~$\half$ as 
indicated by the vanishing of $\gamma(g)$ and $\gamma_m(g)$ for this value.

In the computations deriving (\ref{threerge}) the main strategy was to 
dimensionally regularize (\ref{baregn}) in $d$-dimensions and determine the
renormalization constants as poles in the deviation from two dimensions. Here
we will take $d$~$=$~$2$~$-$~$\epsilon$ with $\epsilon$ being regarded as
small. Whilst the correct renormalization group functions emerged at three 
loops, \cite{8} overlooked a novel feature of the dimensionally regularized 
Lagrangian which was explicitly discussed in \cite{15,16} after the 
observation in \cite{13,14}. Basically (\ref{baregn}) ceases being 
multiplicatively renormalizable in $d$-dimensions but crucially retains 
renormalizability. This is not a property solely restricted to the Gross-Neveu 
model but is a feature of any two dimensional model with a $4$-fermi 
interaction such as the abelian and non-abelian Thirring models, \cite{18}. At
a certain loop order, which is different for different models, evanescent 
operators are generated through the renormalization which are non-trivial in 
$d$-dimensions but which are absent or evaporate in the limit to strictly two 
dimensions which corresponds to the lifting of the regularization. A 
comprehensive study of this problem was provided for general $4$-fermi theories
in \cite{13,14} and we recall those features which are relevant for our 
ultimate goal. The same problem in four dimensions has been considered in 
\cite{19,20}. Though there $4$-fermi operators are of course non-renormalizable
and treated in the context of effective field theories. 

First, in $d$-dimensions the basis of $\gamma$-matrices based on the Clifford
algebra  
\begin{equation}
\{ \gamma^\mu , \gamma^\nu \} ~=~ 2 \eta^{\mu\nu}
\end{equation}
has to be extended to the set of objects $\Gamma_{(n)}^{\mu_1\ldots\mu_n}$ for
integer $n$~$\geq$~$0$, \cite{13,14,15,16,17}, which is totally antisymmetric 
in the Lorentz indices and defined by 
\begin{equation} 
\Gamma_{(n)}^{\mu_1 \mu_2 \ldots \mu_n} ~=~ \gamma^{[\mu_1} \gamma^{\mu_2}
\ldots \gamma^{\mu_n]} 
\end{equation}  
where we use the convention that the square brackets include division by $n!$ 
when all possible permutations of the $\gamma$-strings are written explicitly. 
Then $\Gamma_{(n)}^{\mu_1\ldots\mu_n}$ form a complete closed basis for 
$\gamma$-matrices in $d$-dimensions where $\Gamma_{(0)}$ is the unit matrix.
Hence one can immediately see that the most general multiplicatively 
renormalizable $4$-fermi theory using dimensional regularization in 
$d$-dimensions is, \cite{13,14,15,16,17}, 
\begin{equation} 
L ~=~ i \bar{\psi}_0^{i} \partialslash \psi^{i}_0 ~-~ m_0 \bar{\psi}^{i}_0 
\psi^{i}_0 ~+~ \frac{1}{2} \sum_{n=0}^\infty g_{(n) \, 0} \, \bar{\psi}^i_0
\Gamma_{(n)}^{\mu_1\ldots\mu_n} \psi^i_0 \,
\bar{\psi}^i_0 \Gamma_{(n) ~ \mu_1\ldots\mu_n} \psi^i_0 
\label{baregen} 
\end{equation}  
where there is an infinite number of (bare) couplings $g_{(n)\,0}$ with
$g_{(0)}$~$\equiv$~$g$ identified as the original one of the Gross-Neveu model
(\ref{baregn}). Though the Gross-Neveu model strictly will correspond to the 
case where $g_{(1)}$~$=$~$g_{(2)}$~$=$~$0$ as $\Gamma_{(1)}^\mu$ and 
$\Gamma_{(2)}^{\mu\nu}$ are not evanescent. Given (\ref{baregen}), there are
several points of view depending on the problem in hand. If (\ref{baregen}) is
the most general renormalizable theory in $d$-dimensions, then in principle for
the Gross-Neveu model one must begin with (\ref{baregen}) but omit $g_{(1)}$ 
and $g_{(2)}$. This will produce renormalization group functions dependent, in 
principle, on all evanescent couplings. The true renormalization group 
functions of the original theory would eventually emerge from this 
multiplicatively renormalizable theory by setting $g_{(n)}$~$=$~$0$ for 
$n$~$\geq$~$3$ at the end, \cite{13,14,15,16,17}. Clearly this would involve a 
significant amount of calculation much of which would be redundant in the 
production of the final renormalization group functions. From a practical point
of view there is a less laborious route to follow if one abandons the 
insistence on multiplicative renormalizability, \cite{17,18}. Then operators 
such as  
\begin{equation}
{\cal O}_n ~=~ \half \bar{\psi}^i \Gamma_{(n)}^{\mu_1\ldots\mu_n} \psi^i \,
\bar{\psi}^i \Gamma_{(n) ~ \mu_1\ldots\mu_n} \psi^i 
\end{equation}
for $n$~$\geq$~$3$ will be generated with $g_{(0)}$~$\equiv$~$g$ dependent 
coefficients. The problem for this point of view then becomes one of how to 
extract the {\em true} two dimensional renormalization group functions. It 
turns out that a formalism was developed in \cite{13,14} and used in 
\cite{18,10} for this evanescent operator issue. In essence the true 
renormalization group functions are not strictly determined from what we term 
the naive renormalization constants. By these we mean those required to render 
$2$ and $4$-point functions finite. Instead these naive renormalization group 
functions need to be amended by the effect the evanescent operators have on the 
divergence structure in $d$-dimensions relative to two dimensions. In
\cite{13,14} such a projection formula was introduced which involves projection
functions, $\rho^{(k)}(g)$, $\rho^{(k)}_m(g)$ and $C^{(k)}(g)$, where the index
$k$ ranges over the evanescent range $k$~$\geq$~$3$. These functions quantify 
the effect the evanescent operators have on the derivation of the 
renormalization group functions. The derivation of the projection formula is 
given in \cite{14} and applied additionally in \cite{18,10}. We recall that it 
is  
\begin{eqnarray} 
\left. \int d^d x \, \NN [ \OO_k ] \right|_{ g_{(i)} = 0 \, , \, d = 2 } 
&=& \int d^d x \left( \, \rho^{(k)}(g) \NN [ i 
\bar{\psi} \partialslash \psi ~-~ m \bar{\psi}\psi ~+~ 2g \OO_0 ] \right. 
\nonumber \\ 
&& \left. \left. ~~~~~~~~~-~ \rho^{(k)}_m(g) \, \NN[ m \bar{\psi}\psi ] ~+~ 
C^{(k)}(g) \NN [ \OO_0 ] \right) \right|_{ g_{(i)} = 0 \, , \, d = 2 } 
\label{projfm} 
\end{eqnarray}  
where the normal ordering symbol, $\NN$, is included, \cite{13,14,21,22}. The 
relation strictly only has meaning when inserted in a $2$ or $4$-point Green's 
function. In other words one inserts the evanescent operator of the left side 
of (\ref{projfm}) in a Green's function and evaluates it using the naive 
renormalization constants to yield a finite expression. Equally one inserts the
left side of (\ref{projfm}) into the same Green's function to the same loop 
order and renders it finite. Then the coefficients of the perturbative
expansion in the coupling constant $g$ are chosen order by order to render the 
equation consistent at each loop order after one has set $d$~$=$~$2$. This 
procedure is denoted by the restriction $\{g_{(i)}$~$=$~$0$, $d$~$=$~$2\}$ on 
both sides of (\ref{projfm}). Once the explicit projection functions have been 
determined to the appropriate order, then the {\em true} renormalization group 
functions are given by, \cite{13,14},  
\begin{eqnarray} 
\beta(g) &=& \tilde{\beta}(g) ~+~ \sum_{k=3}^\infty C^{(k)}(g) 
\beta_k(g) \nonumber \\  
\gamma(g) &=& \tilde{\gamma}(g) ~+~ \sum_{k=3}^\infty \rho^{(k)}(g) 
\beta_k(g) \nonumber \\  
\gamma_m(g) &=& \tilde{\gamma}_m(g) ~+~ \sum_{k=3}^\infty \rho^{(k)}_m(g) 
\beta_k(g) 
\label{truerge}
\end{eqnarray} 
where $\tilde{}$ denotes the {\em naive} renormalization group functions. 

For the Gross-Neveu model the first appearance of an evanescent operator is
at three loops which was originally observed in \cite{14,17}. Whilst this
postdates the three loop $\MSbar$ $\beta$-functions of \cite{7,8} the latter 
are unaffected by the generation of ${\cal O}_3$ since it occurs with a 
coupling dependence of $g^3$. So that coupled with $C^{(3)}(g)$ it will only
affect the $\beta$-function itself at four loops. Equally the mass anomalous
dimension of \cite{11} does not feel this evanescent operator presence until
four loops either. We refrain from quoting the value of the associated
$\beta$-function, $\beta_3(g)$, until later. This is primarily because there 
are two completing values given in \cite{17} and \cite{18}. In the former the 
renormalization was deduced in a massless version of (\ref{baregn}) where is 
was claimed that only ladder style diagrams were the origin of ${\cal O}_3$. In
that instance the nullification of two external momenta in the associated 
$4$-point function should not have resulted in spurious infrared singularities.
Whilst a $\beta_3(g)$ was determined, it involved $\zeta(3)$ which was not 
found in \cite{18} which used the massive version, (\ref{baregn}), where 
$\zeta(x)$ is the Riemann zeta function. This clearly avoided infrared 
singularities when all the external momenta were nullified in the $4$-point 
function. Moreover it was claimed that the diagrams leading to ${\cal O}_3$ 
were akin to those analysed in \cite{17} but with no $\zeta(3)$ appearing in 
the published value of $\beta_3(g)$. Though both calculations agreed on the 
rational part of $\beta_3(g)$. The discrepancy between both computations needs 
to be resolved and a four loop calculation which requires $\beta_3(g)$ 
explicitly to obtain the true renormalization group function will provide a 
non-trivial forum in which to achieve this. The correct expression for
$\beta_3(g)$ will be crucial for the four loop $\beta$-function. Given this 
structure of the Gross-Neveu model we can now write down the renormalized form 
of (\ref{baregn}) we will use. It is, \cite{17,18}, 
\begin{equation} 
L ~=~ i Z_\psi \bar{\psi}^{i} \partialslash \psi^{i} ~-~ m Z_\psi Z_m 
\bar{\psi}^{i} \psi^{i} ~+~ \frac{1}{2} g \mu^\epsilon Z_g Z^2_\psi 
( \bar{\psi}^i \psi^i )^2 ~+~ \frac{1}{2} g \mu^\epsilon Z_{33} Z^2_\psi 
\left( \bar{\psi}^i \Gamma_{(3)} \psi^i \right)^2 
\label{rengn}
\end{equation}  
where the renormalized quantities are defined from their bare counterparts by
\begin{equation} 
\psi_0 ~=~ \psi Z_\psi^{\half} ~~~,~~~ m_0 ~=~ m Z_m ~~~,~~~ g_0 ~=~ 
g Z_g \mu^\epsilon 
\label{rencon} 
\end{equation} 
in $d$-dimensions and $Z_{33}$ absorbs the infinity associated with the
generation of ${\cal O}_3$ at this order. Unlike $Z_\psi$, $Z_m$ and $Z_g$ its
coupling constant expansion does not commence with unity. Though we stress that
(\ref{rengn}) is valid only for $2$-point calculations to four loops. Only by 
renormalizing the $4$-point function at four loops would the full evanescent 
operator structure at that order emerge. For instance, it is not inconceivable 
given the $\gamma$-matrix structure of the four loop $4$-point function that a 
${\cal O}_4$ evanescent operator will be generated. From these renormalization 
constants the naive renormalization group functions are given by 
\begin{eqnarray} 
\tilde{\gamma}(g) &=& \mu \frac{\partial}{\partial \mu} \ln Z_\psi ~~~,~~~ 
\tilde{\gamma}_m(g) ~=~ -~ \tilde{\beta}(g) \frac{\partial}{\partial g} \ln Z_m
\nonumber \\
\tilde{\beta}(g) &=& (d-2) g ~-~ g \tilde{\beta}(g) 
\frac{\partial}{\partial g} \ln Z_g 
\end{eqnarray} 
to the order we are working to. For $\beta_3(g)$ one deduces its explicit form
from the simple pole in $\epsilon$ via standard methods, \cite{14}. Thus in the context of (\ref{truerge}) and these observations, we note that for the mass 
anomalous dimension the result of \cite{18} for $\rho_m^{(3)}(g)$ is
\begin{equation}
\rho_m^{(3)}(g) ~=~ -~ \frac{g}{\pi} ~+~ O(g^2) ~.
\end{equation}  
The higher terms are not required since the first term of $\beta_3(g)$ is 
$O(g^3)$.  

\sect{Three loop calculations.}

We begin this section by discussing the computational strategy. To determine 
the mass anomalous dimension for (\ref{baregn}) we consider the $2$-point 
function for the massive theory. In \cite{9} the four loop $\MSbar$ anomalous 
dimension was calculated and independently verified in \cite{10}. Therefore, we
assume that result for $Z_\psi$. However, this is effectively a three orders 
calculation since the one loop snail of Figure $1$ corresponding to 
$\langle \psi_\alpha(p) \bar{\psi}^\beta(-p) \rangle$ has no non-zero 
contributions in the $\pslash$ channel for the massive or massless Lagrangians 
where $p$ is the external momentum. Moreover, since for this case one is 
interested only in $Z_\psi$, it sufficed to consider the {\em massless} theory 
whence one only has to determine massless Feynman integrals. The component 
involving $\pslash_\alpha^{~\beta}$ can be deduced by multiplying all diagrams 
by $\pslash$ and taking the spinor trace. For the mass dimension one cannot 
follow this strategy. Not only because the one loop diagram contributes but 
also because its contribution to $Z_m$ requires the presence of the mass 
itself. Therefore unlike the determination of $Z_\psi$ one cannot neglect the 
snail graph of Figure $1$ at one loop as well as the graphs where snails appear
as subgraphs at higher order. However, given that one is only interested in the
$m \delta_\alpha^{~\beta}$ channel of $\langle \psi_\alpha(p) 
\bar{\psi}^\beta(-p) \rangle$ the Green's function can be analysed by 
nullifying the external momentum. Taking the spinor trace produces Lorentz 
scalar integrals but with tensor structure resulting from internal momenta 
contractions. The presence of the common mass $m$ automatically protects 
against the appearance of spurious infrared infinities and relegates the 
problem of determining the ultraviolet structure to mapping the integrals with 
internal momenta contractions to a set of basic master vacuum bubbles at each 
loop order. The problem of studying massive vacuum bubbles in {\em four}
dimensions has received much attention over the years, culminating in, for 
example, the {\sc Matad} package at three loops, \cite{24}, and the
comprehensive study by Broadhurst of all combinations of massive and massless 
propagators in the Benz or tetrahedron topology, \cite{25}. The analogous 
problem relative to two dimensions has not been treated as systematically. 
Though the main results to three loops have appeared within various articles. 
Additionally, at four loops we will have to handle new integrals for topologies
which do not simply break into products of lower loop vacuum bubbles. The main 
difficulty lies in having to handle the tensor structure emanating from the 
fermion propagator. Throughout we have made extensive use of the symbolic 
manipulation language {\sc Form}, \cite{12}, in which to code our algorithm 
where the contributing Feynman diagrams to the $2$ and $4$-point functions are 
generated automatically with the {\sc Qgraf} package, \cite{23}. To summarize 
there are $1$ one loop, $2$ two loop, $7$ three loop and $36$ four loop graphs 
for the $2$-point function. For the $4$-point function there are $3$ one loop, 
$18$ two loop and $138$ three loop graphs to renormalize. 

\vspace{0.5cm} 
\begin{figure}[hb]  
\hspace{6.5cm} 
\epsfig{file=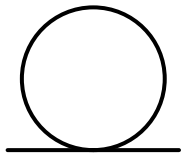,height=2cm} 
\vspace{0.5cm} 
\caption{One loop contribution to the $2$-point function.} 
\end{figure} 

At three loops we now summarize the few master (massive) integrals which will 
be of interest to us, in our notation and conventions. First, Figure $2$ 
denotes the basic vacuum bubbles to two loops. We define the first graph of 
Figure $2$ as  
\begin{equation} 
I ~=~ i \int_k \frac{1}{[k^2-m^2]} ~. 
\end{equation} 
We work in Minkowski space and choose to include a factor of $i$ with each
integration measure which is abbreviated by
\begin{equation}
\int_k ~=~ \int \frac{d^dk}{(2\pi)^d} ~. 
\end{equation}
The integral $I$ is trivial to deduce from the Euler $\beta$-function as 
\begin{equation} 
I ~=~ \frac{\Gamma(1-d/2)}{(4\pi)^{d/2}} (m^2)^{d/2-1} ~. 
\end{equation} 
Hence the middle graph of Figure $2$ is $I^2$. 

\vspace{0.5cm} 
\begin{figure}[hb]  
\hspace{3cm} 
\epsfig{file=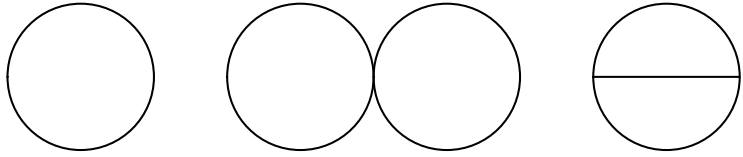,height=2cm} 
\vspace{0.5cm} 
\caption{One and two loop vacuum bubbles.} 
\end{figure} 

The final graph we denote by $\Delta(0)$, \cite{11}, where 
\begin{equation} 
\Delta(0) ~=~ i \int_k \frac{J(k^2)}{[k^2-m^2]} 
\end{equation} 
and 
\begin{equation} 
J(p^2) ~=~ i \int_k \frac{1}{[k^2-m^2][(k-p)^2 - m^2]} 
\label{jdef} 
\end{equation} 
is the basic one loop self-energy bubble. There is a sequence of integrals
related to $J(p)$ defined by 
\begin{equation} 
J_{\alpha\beta}(p) ~=~ i \int_k 
\frac{1}{[k^2-m^2]^\alpha[(k-p)^2 - m^2]^\beta} 
\end{equation} 
where we choose $J_{21}(p)$~$\equiv$~$K(p^2)$. In this form a Gauss relation
of the hypergeometric functions gives the relation  
\begin{equation}
(p^2-4m^2) K(p^2) ~=~ J(0) ~-~ (d-3) J(p^2)
\label{cut1}
\end{equation}
with
\begin{equation}
J(0) ~=~ -~ \frac{\Gamma(2-d/2)}{(4\pi)^{d/2}} (m^2)^{d/2-2} 
\end{equation} 
since explicit calculations produce 
\begin{equation}
J(p^2) ~=~ -~ \frac{\Gamma(2-d/2)}{(4\pi)^{d/2}} \left( \frac{4m^2-p^2}{4}
\right)^{d/2-2} \, {}_2F_1 \left( 2 - \frac{d}{2}, \frac{1}{2}; \frac{3}{2};
\frac{p^2}{p^2-4m^2} \right)
\end{equation} 
and 
\begin{equation}
K(p^2) ~=~ \frac{\Gamma(3-d/2)}{2(4\pi)^{d/2}} \left( \frac{4m^2-p^2}{4}
\right)^{d/2-3} \, {}_2F_1 \left( 3 - \frac{d}{2}, \frac{1}{2}; \frac{3}{2};
\frac{p^2}{p^2-4m^2} \right) ~. 
\end{equation} 
Likewise, at the subsequent level 
\begin{equation}
(p^2-4m^2) \left( J_{22}(p^2) ~+~ 2 J_{31}(p^2) \right) ~=~ 2 K(0) ~-~ 2 (d-5) 
K(p^2)
\label{cut2}
\end{equation}
whence 
\begin{eqnarray}
J_{31}(p^2) &=& \frac{(d-6)}{2p^2} K(p^2) ~+~ \frac{(p^2-2m^2)}{p^2(p^2-4m^2)}
\left( K(0) - (d-5) K(p^2) \right) \nonumber \\ 
J_{22}(p^2) &=& -~ \frac{(d-6)}{p^2} K(p^2) ~+~ \frac{4m^2}{p^2(p^2-4m^2)}
\left( K(0) - (d-5) K(p^2) \right) ~.  
\label{cut3}
\end{eqnarray}
These rules are used extensively for the two and higher loop Feynman integrals.
The integral $\Delta(0)$ is finite in two dimensions and can be evaluated in an
expansion in powers of $\epsilon$ as
\begin{equation}
\Delta(0) ~=~ -~ \frac{9s_2}{16\pi^2m^2} ~+~ O(\epsilon)
\label{del0}
\end{equation}
where $s_2$~$=$~$(2\sqrt{3}/9) \mbox{Cl}_2(2\pi/3)$ with $\mbox{Cl}_2(x)$ the
Clausen function. The analogous four dimensional vacuum bubble also contains 
$s_2$ in its finite part but is divergent. In principle the $O(\epsilon)$ term 
of (\ref{del0}) can be deduced. However, throughout our computations we left 
$\Delta(0)$ itself unevaluated since on renormalizability grounds it must be 
absent from the final renormalization constants at higher loops. This is 
because if the $2$-point function did not have its external momenta nullified 
then the integral $\Delta(p)$ would emerge, where 
\begin{equation} 
\Delta(p) ~=~ i \int_k \frac{J(k^2)}{[(k-p)^2 - m^2]} ~. 
\end{equation} 
Clearly such a non-local function of the external momenta could not be
retained when all the counterterms are included. 

\vspace{0.5cm} 
\begin{figure}[hb]  
\epsfig{file=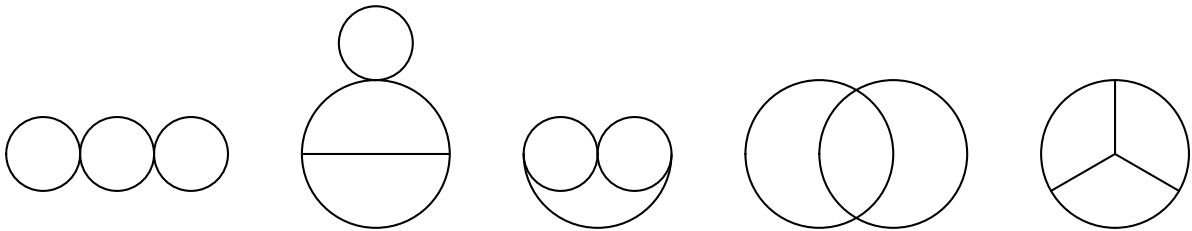,height=3cm} 
\vspace{0.5cm} 
\caption{Three loop vacuum bubbles.} 
\end{figure} 

At three loops there are several more basic topologies. If one ignores for the
moment the complication due to the presence of internal momenta contractions
in integral numerators, then the basic graphs are given in Figure $3$. The
first involves $I$ and its derivatives with respect to $m^2$. Also the second
is a variation on $\Delta(0)$ and we note that the two loop subgraph is  
\begin{equation} 
i \int_k \frac{J(k^2)}{[k^2-m^2]^2} ~=~ \frac{(d-3)}{3m^2} \Delta(0)
\end{equation} 
which is established by differentiating $\Delta(0)$ with respect to $m^2$.
Aside from the Benz topology the remaining vacuum bubbles of Figure $3$ are
related to the integrals $i\int_k J^2(k^2)/[k^2-m^2]$, $i\int_k J^2(k^2)$ and 
$i\int_k J(k^2) K(k^2)$. Similar to $\Delta(0)$ these are finite in two
dimensions but their values are required at four loops when multiplied by
counterterms. As only the leading term in $\epsilon$ is required in each case
it transpires that we can set $d$~$=$~$2$ and use the fact that in Euclidean
space, denoted by the subscript $E$,
\begin{equation}
\left. J_E(-k^2) \right|_{d=2} ~=~ \left. \frac{\theta}{\sinh\theta}  J(0) 
\right|_{d=2}
\end{equation}
upon the change of variables $k^2$~$=$~$4m^2\sinh^2(\theta/2)$ where
$\left. J(0) \right|_{d=2}$~$=$~$-$~$1/(4\pi m^2)$. Then, for instance,  
\begin{equation}
i \int_k J^2(k^2) ~=~ -~ \frac{m^2}{2\pi} \left. J^2(0) \right|_{d=2} 
\int_0^\infty d \theta \, \frac{\theta^2}{\sinh\theta} ~+~ O(\epsilon) ~. 
\end{equation} 
This can be evaluated from standard integrals to give 
\begin{equation}
i \int_k J^2(k^2) ~=~ -~ \frac{7\zeta(3)}{64\pi^3m^2} ~+~ O(\epsilon) ~. 
\label{j2def} 
\end{equation} 
Equally we find 
\begin{equation}
i \int_k \frac{J^2(k^2)}{[k^2-m^2]} ~=~ \frac{11\zeta(3)}{576\pi^3m^4} ~+~ 
O(\epsilon) ~. 
\label{j21def} 
\end{equation} 
Though in $d$-dimensions one can derive the relation
\begin{equation}
i \int_k J(k^2) K(k^2) ~=~ \frac{(3d-8)}{8m^2} \, i \int_k J^2(k^2) ~.
\end{equation} 
The presence of $K(k^2)$ in several of the master integrals with topology 
similar to those of Figure $3$ produces similar finite integrals whose finite
part is required and which is determined in an analogous way. We note that 
\begin{eqnarray}
i \int_k \frac{J(k^2)}{[k^2-4m^2]} &=& \left. \frac{\ln(2)}{2\pi} J(0) 
\right|_{d=2} ~+~ O(\epsilon)
\nonumber \\  
i \int_k \frac{J^2(k^2)}{[k^2-4m^2]} &=& \left[ \frac{7}{8} \zeta(3) ~-~
\ln(2) \right] \left. \frac{J^2(0)}{2\pi} \right|_{d=2} ~+~ O(\epsilon) ~.
\end{eqnarray} 
For the mass anomalous dimension these basic vacuum bubbles suffice to 
determine the renormalization constants to three loops. Given the nature of the
$4$-point interaction in (\ref{baregn}) the Benz topology does not occur in the
$2$-point function at this loop order.

Having discussed the basic scalar master integrals which result we briefly
note the algorithm dealing with the numerator structure of the integrals.
This has been systematically quantified in \cite{11}. However, we note that
repeated use of
\begin{equation}
kl ~=~ \frac{1}{2} \left[ k^2 ~+~ l^2 ~-~ [(k-l)^2-m^2] ~-~ m^2 \right]
\label{propdec1} 
\end{equation}
and then
\begin{equation}
k^2 ~=~ [k^2-m^2] ~+~ m^2 ~~~,~~~ l^2 ~=~ [l^2-m^2] ~+~ m^2
\label{propdec2} 
\end{equation}
in each contributing topology where there are $[k^2-m^2]$, $[l^2-m^2]$ and
$[(k-l)^2-m^2]$ propagators already. This is done in such a way that powers of
$kl$ can remain when all mixed $[(k-l)^2-m^2]$ propagators are absent and one 
does not then continue substituting for $kl$. In such integrals one can use 
Lorentz symmetry in the $k$ and $l$ subgraph integrals to redefine even powers 
of $kl$ as proportional to $k^2 l^2$ or zero if there are an odd number of 
factors of $kl$. Then (\ref{propdec2}) is repeated. Consequently several 
variations in the basic bubble graphs of Figure $3$ emerge and we note that  
\begin{eqnarray}
i \int_k k^2 J^2(k^2) &=& \frac{4}{3} I^3 ~+~ \frac{4}{3} m^2 i \int_k J^2(k^2) 
\nonumber \\
i \int_k (k^2)^2 J^2(k^2) &=& \frac{8m^2}{3(3d-4)} \left[ (5d-6) I^3 ~+~ 
2d m^2 \, i \int_k J^2(k^2) \right] 
\end{eqnarray} 
where these are exact and no finite parts have been omitted since these are
crucial for the next loop order. In essence this summarizes the key ingredients
in the algorithm for evaluating the three loop mass anomalous dimension which 
has been coded in {\sc Form} and reproduces the previous three loop $\MSbar$ 
Gross-Neveu mass anomalous dimension.

\sect{Three loop $4$-point function renormalization.} 

At this point we turn to our secondary aim which is to resolve the discrepancy 
in the renormalization associated with the generation of ${\cal O}_3$. This 
requires the complete determination of the $4$-point function divergence 
structure at three loops. Whilst the algorithm to do this is very similar to 
that of the $2$-point function there are several key differences. First, the 
$4$-point function divergences will be independent of the external momenta 
which means that they can be immediately nullified. The mass again protects 
against spurious infrared divergences. However, we cannot now take the Lorentz 
traces since that would prevent one from seeing the emergence of any 
$\Gamma_{(3)}^{\mu\nu\sigma} \otimes \Gamma_{(3) ~ \mu\nu\sigma}$ 
$\gamma$-matrix structure. Instead we have to retain $\gamma$-strings and also 
unentangle the internal momenta within them. Hence one decouples the Feynman 
diagrams into $\gamma$-strings and Lorentz tensor vacuum bubbles. At one and 
two loops the resulting tensor integrals for the whole integral can be 
straightforwardly reduced by noting that at one loop 
\begin{equation}
\int_k k^\mu k^\nu f_1(k^2) ~=~ \frac{\eta^{\mu\nu}}{d} \int_k k^2 f_1(k^2)
\end{equation} 
where $k$ is the sole loop momentum and at two loops 
\begin{eqnarray}
\int_{kl} k_1^{\mu_1} k_2^{\mu_2} k_3^{\mu_3} k_4^{\mu_4} 
f_2(k,l) &=& \frac{1}{d(d-1)(d+2)} \nonumber \\
&& \int_{kl} \! \! f_2(k,l) \left[  
\left[ (d+1) k_1.k_2 k_3.k_4 - k_1.k_3 k_2.k_4 - k_1.k_4 k_2.k_3 \right]
\eta^{\mu_1\mu_2} \eta^{\mu_3\mu_4} \right. \nonumber \\
&& \left. ~~~~~~~~+  
\left[ (d+1) k_1.k_3 k_2.k_4 - k_1.k_2 k_3.k_4 - k_1.k_4 k_2.k_3 \right]
\eta^{\mu_1\mu_3} \eta^{\mu_2\mu_4} \right. \nonumber \\
&& \left. ~~~~~~~~+  
\left[ (d+1) k_1.k_4 k_2.k_3 - k_1.k_2 k_3.k_4 - k_1.k_3 k_2.k_4 \right]
\eta^{\mu_1\mu_4} \eta^{\mu_2\mu_3} \right] \nonumber \\ 
\end{eqnarray}
where $k_i$~$\in$~$\{k,l\}$ and in the Lorentz tensor of the integrand all
possible combinations of the two internal momenta are covered. The functions
$f_i(\{k_i\})$ represent the various possible propagator combinations. For
clarity we have included the dot of the scalar products explicitly. At three 
loops the situation is complicated by the observation that the extension of 
both these formula gives 
\begin{eqnarray}
\int_{klq} k_1^{\mu_1} k_2^{\mu_2} k_3^{\mu_3} k_4^{\mu_4} k_5^{\mu_5} 
k_6^{\mu_6} f_3(k,l,q) 
&=& \frac{\eta^{\mu_1\mu_2} \eta^{\mu_3\mu_4} \eta^{\mu_5\mu_6}}
         {d(d-1)(d-2)(d+2)(d+4)} \nonumber \\
&& \int_{klq} f_3(k,l,q) \left[  (d^2+3 d-2) k_1.k_2 k_3.k_4 k_5.k_6   
            \right. \nonumber \\
&& \left. ~~  
           - (d+2) k_1.k_2 k_3.k_5 k_6.k_4  
            - (d+2) k_1.k_2 k_3.k_6 k_4.k_5   
            \right. \nonumber \\
&& \left. ~~  
            - (d+2) k_1.k_3 k_2.k_4 k_5.k_6  
            + 2 k_1.k_3 k_2.k_5 k_6.k_4   
            \right. \nonumber \\
&& \left. ~~ 
            + 2 k_1.k_3 k_2.k_6 k_4.k_5 
            - (d+2) k_1.k_4 k_2.k_3 k_5.k_6   
            \right. \nonumber \\
&& \left. ~~ 
            + 2 k_1.k_4 k_2.k_5 k_6.k_3  
            + 2 k_1.k_4 k_2.k_6 k_3.k_5  
            \right. \nonumber \\
&& \left. ~~ 
            + 2 k_1.k_5 k_2.k_3 k_4.k_6  
            + 2 k_1.k_5 k_2.k_4 k_6.k_3   
            \right. \nonumber \\
&& \left. ~~ 
            - (d+2) k_1.k_5 k_2.k_6 k_3.k_4 
            + 2 k_1.k_6 k_2.k_3 k_4.k_5   
            \right. \nonumber \\
&& \left. ~~ 
            + 2 k_1.k_6 k_2.k_4 k_5.k_3  
            - (d+2) k_1.k_6 k_2.k_5 k_3.k_4 \right] 
            \nonumber \\
&& ~+~ 
\mbox{$14$ similar terms}
\label{tensdec3}
\end{eqnarray} 
where $k_i$~$\in$~$\{k,l,q\}$. The full decomposition is clearly quite large.
However, it is the appearance of the $1/(d-2)$ factor which is novel. In 
\cite{17,18} the full set of three loop graphs in both massless and massive
cases where a divergent $\Gamma_{(3)}^{\mu\nu\sigma} \otimes \Gamma_{(3) ~ 
\mu\nu\sigma}$ structure emerged, was noted. The sets of graphs appear to be
the same. Though in \cite{17} it is not fully clear which the actual ladder 
graphs referred to are. However, the seemingly {\em finite} graph of Figure $4$
was regarded as fully finite in {\em all} $\gamma$-string channels, \cite{18}. 
In our present reconsideration it transpires that within the integral of the 
graph of Figure $4$ there is a divergent contribution to the
$\Gamma_{(3)}^{\mu\nu\sigma} \otimes \Gamma_{(3) ~ \mu\nu\sigma}$ channel but
{\em not} for  the $\Gamma_{(0)} \otimes \Gamma_{(0)}$ one. This derives from 
the pole $1/(d-2)$ in (\ref{tensdec3}) producing the massive Benz integral 
corresponding to the final graph of Figure $3$. The key part is then 
\begin{equation}
\frac{1}{(d-2)}
\int_{klq} \frac{1}{[k^2-m^2] [l^2-m^2] [q^2-m^2] [(k-l)^2-m^2] [(k-q)^2-m^2]
[(l-q)^2-m^2]} 
\label{betabenz}
\end{equation}
where the actual integral itself is finite in two dimensions. It remains after 
repeated application of (\ref{propdec1}) and (\ref{propdec2}) in the scalar 
integrals of (\ref{tensdec3}). However, to have the complete divergence
structure the integral needs to be evaluated since it will contribute to
$Z_{33}$. The remaining integrals with this $1/(d-2)$ pole in the 
$\Gamma_{(3)}^{\mu\nu\sigma} \otimes \Gamma_{(3) ~ \mu\nu\sigma}$ channel 
correspond to (\ref{betabenz}) but with one or more propagators removed after
application of (\ref{propdec1}) and (\ref{propdec2}). These can be evaluated
from the three loop techniques discussed earlier. In \cite{18} this 
contribution, (\ref{betabenz}), was overlooked since it was assumed that the 
parent integral with the internal momenta contracted was finite without noting 
the possibility of the $1/(d-2)$ factor deriving from the tensor decomposition.
In relation to \cite{17} we can only comment that in the massless version of 
(\ref{betabenz}) the integral will be zero. However, given the totally 
different method of calculating the $4$-point function of \cite{18} in the 
massless case, a contribution analogous to (\ref{betabenz}) could possibly 
arise elsewhere. 

\vspace{0.5cm} 
\begin{figure}[ht] 
\hspace{6.5cm} 
\epsfig{file=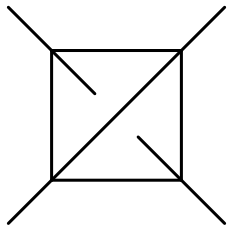,height=3cm} 
\vspace{0.5cm} 
\caption{Three loop contribution to $4$-point function.}
\end{figure} 

There remains the task now of evaluating the {\em integral} of 
(\ref{betabenz}). Although finite it clearly cannot be reduced to any of the 
three loop master vacuum bubbles already discussed even using, say, integration
by parts. Instead we have had to resort to the more extensive experience of 
four dimensional vacuum bubble diagrams and specfically the Benz graphs 
discussed in \cite{25}. To promote (\ref{betabenz}) to four dimensions we 
exploit Tarasov's observation of relating $d$-dimensional integrals to 
$(d+2)$-dimensional ones, \cite{26,27}. Moreover, this is straightforward to do
via the {\sc Tarcer} package, \cite{28}, written in {\sc Mathematica} for the 
basic two loop self energy topology given in Figure $5$. Specifically one 
feature of {\sc Tarcer} is that one can relate this two loop self energy graph 
in $d$-dimensions to that in $(d+2)$-dimensions. This is a subgraph of Figure 
$4$ with nullified external momenta and given that this is a three loop vacuum 
bubble, the final three loop integration measure can be rewritten as
\begin{equation}
\int \frac{d^dk}{(2\pi)^d} ~=~ 2 \pi d \int \frac{d^{d+2}k}{(2\pi)^{d+2}} 
\frac{1}{k^2} 
\label{intmes}
\end{equation}
in our conventions since the two loop subgraph will clearly be a function of
$k^2$ only. From (\ref{intmes}) a massless propagator will appear in the higher 
dimensional integral. Since all the lines of Figures $4$ and $5$ are massive 
and both final integrations involve functions of the square of the momentum, 
then we find the relation between the $d$-dimensional massive Benz graph and 
similar topologies in two dimensions higher is
\begin{eqnarray}
&& \mbox{Be}(1,1,1,1,1,1,m^2,m^2,m^2,m^2,m^2,m^2,d) \nonumber \\
&& = -~ \frac{1}{12m^4} i \int_k J^2(k^2) ~-~ \frac{3}{4m^2}
i \int_k \frac{J^2(k^2)}{[k^2-m^2]} \nonumber \\
&& ~~~~+~ \frac{\pi d(d-1)(d-2)}{m^6} \left[ 
\mbox{Be}(1,1,1,1,1,1,m^2,m^2,m^2,m^2,m^2,m^2,d+2) \right. \nonumber \\
&& \left. ~~~~~~~~~~~~~~~~~~~~~~~~~~~~~~~~~-~ 
\mbox{Be}(1,1,1,1,1,1,0,m^2,m^2,m^2,m^2,m^2,d+2) \right] 
\label{dimreln}
\end{eqnarray}
where we define  
\begin{eqnarray} 
&& \mbox{Be}(\alpha,\beta,\gamma,\rho,\lambda,\theta,m_1^2,m_2^2,m_3^2,
m_4^2,m_5^2,m_6^2,d) \nonumber \\
&& =~ i^3 \int_{klq} \frac{1}{[k^2-m_1^2]^\alpha [l^2-m_2^2]^\beta 
[q^2-m_3^2]^\gamma [(k-l)^2-m_4^2]^\rho [(k-q)^2-m_5^2]^\lambda 
[(l-q)^2-m_6^2]^\theta} \nonumber \\  
\end{eqnarray}
and emphasise that $\int_k$ indicates a $d$-dimensional integration. The key 
part is the piece which represents the difference in two Benz topologies in 
$(d+2)$-dimensions where one is completely massive and the other has one
massless line. However, since these are multiplied by $(d-2)$ then in our 
$\epsilon$ expansion relative to two dimensions we note that the leading term 
of each is $O(\epsilon)$ meaning that 
\begin{eqnarray}
&& i^3 \int_{klq} \frac{1}{[k^2-m^2] [l^2-m^2] [q^2-m^2] [(k-l)^2-m^2] 
[(k-q)^2-m^2] [(l-q)^2-m^2]} \nonumber \\  
&& ~~~=~ -~ \frac{\zeta(3)}{192\pi^3m^6} ~+~ O(\epsilon) 
\end{eqnarray}
from (\ref{j2def}) and (\ref{j21def}). This is because whilst each of the two 
$(d+2)$-dimensional integrals are divergent in four dimensions due to the 
presence of a simple pole in the regularizing parameter, the difference in 
(\ref{dimreln}) is finite and the residue is independent of the masses in 
either Benz topology, \cite{25}. 

\vspace{0.5cm} 
\begin{figure}[ht]
\hspace{6.5cm} 
\epsfig{file=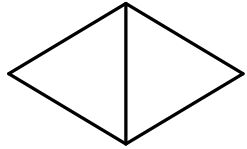,height=2cm} 
\vspace{0.5cm} 
\caption{Two loop self-energy topology central to {\sc Tarcer}.}
\end{figure} 

With this observation all the ingredients are assembled to repeat the full 
three loop renormalization of the $4$-point function of (\ref{baregn}). In
\cite{7} only the $\Gamma_{(0)} \otimes \Gamma_{(0)}$ part was isolated but 
this was sufficient to deduce the $\beta$-function at three loops. It is 
satisfying to record that we have verified the previous three loop $\MSbar$ 
result of \cite{7,8}. However, by contrast we find that a different 
renormalization constant from \cite{17} and \cite{18} emerges for $Z_{33}$. We 
find  
\begin{equation} 
Z_{33} ~=~ \left[ \frac{\zeta(3)}{64} - \frac{1}{48} \right] 
\frac{g^3}{\pi^3\epsilon} ~+~ O(g^4)  
\end{equation} 
whence 
\begin{equation}
\beta_3(g) ~=~ \left[ \frac{3\zeta(3)}{64} ~-~ \frac{1}{16} \right] 
\frac{g^3}{\pi^3} ~+~ O(g^4) ~. 
\label{beta3}
\end{equation}  
Though there is universal agreement on the rational part of (\ref{beta3}),
\cite{17,18}, only the contribution from the diagram of Figure $5$ to the 
$\Gamma_{(3)}^{\mu\nu\sigma} \otimes \Gamma_{(3) ~ \mu\nu\sigma}$ channel
produces the irrational piece thereby confirming the overall structure
observed in \cite{17}. However, rather than finding that we produce one of the
previous values for $Z_{33}$ we are in the seemingly unfortunate position of
finding a new alternative. To determine which is correct and consistent it will
transpire that the four loop mass anomalous dimension is the correct testing 
ground for this in the context of (\ref{truerge}).

\sect{Four loop vacuum bubbles.}

In this section we return to out initial aim and summarize the evaluation of 
the underlying four loop vacuum bubbles required for the mass anomalous 
dimension. For the $2$-point function there are $18$ distinct topologies and 
$36$ Feynman diagrams to be considered. Of these topologies $14$ involve snail 
insertions in one way or another and hence their determination is in effect 
relegated to the straightforward extension of the three loop topology 
discussion. One effect of a snail is to produce two propagators on a line of a 
three loop graph but this can be reproduced by differentiating that line with 
respect to $m^2$. This is also a reason why the three loop vacuum bubbles were 
required to be evaluated to the finite part exactly or left in terms of 
$\Delta(0)$ and other known integrals whose $\epsilon$ expansion could be
substituted when required, if at all. Several topologies contributing to the 
$2$-point function, however, have a more demanding evaluation. These are 
illustrated in Figure $6$ and we concentrate on these for the main part. 
Essentially the main complication now derives from rewriting the scalar 
products of internal momenta in terms of the propagator structure. For all the 
integrals which result we used several interconnected techniques. 

\vspace{0.5cm} 
\begin{figure}[ht] 
\hspace{2.5cm} 
\epsfig{file=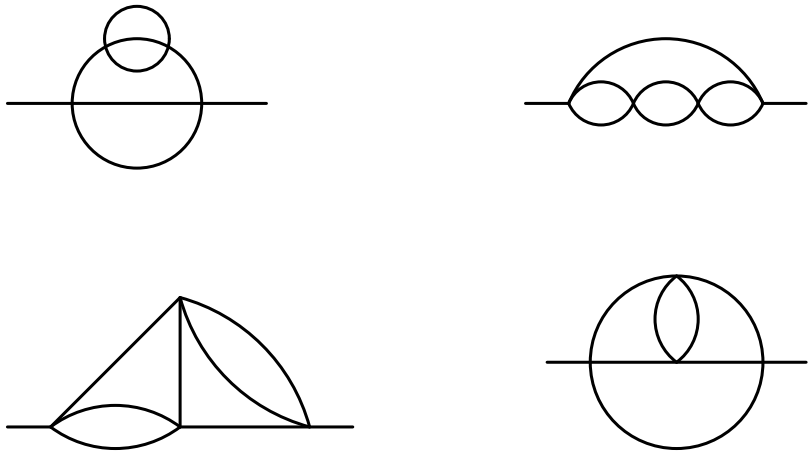,height=6cm} 
\vspace{0.5cm} 
\caption{Several four loop diagrams contributing to the $2$-point function.}
\end{figure} 

First was the use of the {\sc Tarcer} package, \cite{28}, again, particularly 
for the third and fourth graphs of Figure $6$. Clearly the last graph contains 
the two loop self-energy topology of Figure $5$ as a subgraph and the third has
a similar two loop subgraph but with one line removed. Unlike the properties of 
{\sc Tarcer} we described previously, the feature exploited in this instance is
the ability to relate diagrams with different powers of the propagators in 
Figure $5$ to that with unit power. Further, {\sc Tarcer} reduces integrals
involving powers of the scalar products $kl$, $kp$ and $lp$ where $k$ and $l$ 
are internal and $p$ is the external momentum in Figure $5$. The point is that 
the Lorentz tensor reduction for these situations can only be performed by this
route. Any one loop subgraph of Figure $5$ will involve three external legs and
the invariant decomposition in this case is too intricate. By contrast where 
possible we did exploit the Lorentz structure of subgraphs with {\em one} 
internal momentum flowing through it which can be regarded as a $2$-point 
function external momentum for that subgraph. Then integrals can be rewritten 
using results such as 
\begin{eqnarray}
i \int_l \frac{l^\mu l^\nu}{[l^2-m^2][(k-l)^2-m^2]} &=& \frac{1}{(d-1)} 
i \int_l \frac{1}{[l^2-m^2][(k-l)^2-m^2]} \nonumber \\
&& ~~~~~~~~~~~~~\times \! \left[ \eta^{\mu\nu} \left( l^2
- \frac{kl^2}{k^2} \right) - \frac{k^\mu k^\nu}{k^2} \left( l^2
- d \frac{kl^2}{k^2} \right) \right] . 
\label{lordec} 
\end{eqnarray} 
The outcome of the {\sc Tarcer} implementation is to reduce these more
complicated tensor integrals to a set of master scalar four loop vacuum bubbles
since the resulting combination of internal momenta allows for the repeated 
application of (\ref{propdec1}) and (\ref{propdec2}). 

The use of (\ref{lordec}) and {\sc Tarcer} though may appear to introduce 
potential infrared difficulties. However, it transpires that in the full sum of
all contributing pieces to a Feynman graph it can be checked that no integral 
retains an unprotected factor of $1/k^2$ which would give an infrared 
divergence upon integrating over the internal momentum $k$. For one instance 
checking this proved to be a tedious non-trivial exercise which we document for 
completeness. In all bar the second graph of Figure $6$ the following 
combination of integrals emerge 
\begin{equation}
V_\Delta ~=~ i \int_k \left[ J(k^2) ~-~ J(0) \right] \frac{\Delta(k)}{k^2} ~. 
\end{equation}
Clearly each could be infrared divergent but the above combination always 
appears. Defining  
\begin{equation} 
K_\mu(p) ~=~ i \int_k \frac{k_\mu}{[k^2-m^2]^2[(k-p)^2 - m^2]} 
\label{vdeldef}
\end{equation} 
then one can show
\begin{equation}
p^\mu K_\mu(p) ~=~ 2 m^2 K(p^2) ~-~ \frac{1}{2} (d-4) J(p^2) ~=~ 
\frac{1}{2} \left[ p^2 K(p^2) ~+~ J(p^2) ~-~ J(0) \right] ~. 
\end{equation} 
Using this and integration parts in (\ref{vdeldef}) one finds
\begin{equation}
V_\Delta ~=~ (d-3) i \int_k \frac{1}{k^2} J(k^2) \Delta(k^2) ~-~ 2 i \int_k
\Delta(k) K(k^2) ~+~ i^2 \int_{kl} 
\frac{(k^2-m^2) J(k^2) J(l^2)}{l^2[(k-l)^2-m^2]^2} ~. 
\label{vdelman}
\end{equation}
The final integral can be reduced using {\sc Tarcer} if one regards the $k$ 
momentum as external to the self-energy graph of Figure $5$ and $J(l^2)$ is
replaced by the Feynman integral of (\ref{jdef}). Consequently, {\sc Tarcer}
produces 
\begin{eqnarray}  
i \int_{l} \frac{J(l^2)}{l^2[(k-l)^2-m^2]^2} &=&
\frac{(d-2)^2(d-4)I^2}{2(d-3)m^2[k^2-m^2]^2} ~-~ 
\frac{(d-2)(d-3)}{2m^2[k^2-m^2]} i \int_l \frac{1}{l^2[(k-l)^2-m^2]} 
\nonumber \\
&& +~ \frac{(d-4)(3d-8)}{[k^2-m^2]^2} i^2 \int_{lq} 
\frac{1}{[l^2-m^2][(k-q)^2-m^2][(l-q)^2-m^2]} \nonumber \\
&& +~ \frac{[(d-2)[k^2-m^2]-8(d-4)m^2]}{[k^2-m^2]^2} \nonumber \\
&& ~~~~ \times ~ i^2 \int_{lq} \frac{1}{[l^2-m^2]^2 [(k-q)^2-m^2] 
[(l-q)^2-m^2]} ~.
\label{vdeltar} 
\end{eqnarray} 
The benefit of rearranging the two loop integral of the left hand side is to
isolate the potential infrared singularity into a simple term on the right hand
side. Moreover, the appearance of powers of $1/[k^2-m^2]$ will lead to
simplifications when substituted back into the expression for $V_\Delta$ and
the term with the singularity will actually combine with the first term of
(\ref{vdelman}) to produce $V_\Delta$ but with a factor of $(d-3)$. Hence,
evaluating the remaining integrals of (\ref{vdeltar}) in the context of  
(\ref{vdelman}) one arrives at the expression
\begin{eqnarray}
V_\Delta &=& -~ i \int_k \Delta(k) K(k^2) ~-~ 
\frac{(d-2)^2I^2\Delta(0)}{2(d-3)m^2} \nonumber \\
&& -~ (3d-8) i \int_k \frac{J(k^2)\Delta(k)}{[k^2-m^2]} ~+~ 8 m^2 i^2 \int_{kl}
\frac{J(k)K(l)}{[k^2-m^2] [(k-l)^2-m^2]}
\end{eqnarray}
which has no potential infrared singular term. Moreover, each term of the right
side of this is ultraviolet finite in two dimensions. So, in fact, when the
combination $V_\Delta$ appears in our computation, it can actually be dropped
as there is no contribution to the renormalization of the mass at four loops.

Having completed the tensor reduction of the scalar propagators, all that
remains is the evaluation of a set of divergent four loop master integrals akin
to those discussed earlier. Most of these are elementary given the results of
(\ref{cut1}), (\ref{cut2}) and (\ref{cut3}) and the observation that as all 
integrals are infrared finite then one can ignore those four loop ones which 
are clearly ultraviolet finite by the usual counting rules. Though one integral
is worth recording and that is   
\begin{equation}
i \int_k (k^2)^2 J^3(k^2) ~=~ 2 I^4 ~+~ \frac{(7d-13)m^2I}{(2d-5)} \, i
\int_k J^2(k^2) ~+~ \frac{2(d-1)(d-3)}{(2d-5)(d-2)} m^4 \, i \int_k J^3(k^2) 
\end{equation}
because in the determination of this relation a pole in $(d-2)$ emerges in the
standard $d$-dimensional manipulations such as differentiating the original
integral with respect to $m^2$ and using (\ref{cut1}). This pole gives rise to
a problem similar to that discussed for (\ref{betabenz}) but with a simpler
resolution since we merely apply the technique used to deduce (\ref{j2def}) and
(\ref{j21def}), to find 
\begin{equation}
i \int_k J^3(k^2) ~=~ \frac{3\zeta(3)}{256\pi^4m^4} ~+~ O(\epsilon) ~. 
\end{equation} 
The need to evaluate $i\int_k J^2(k^2)$ to the finite part as well is also
illustrated by this equation. 

This completes the discussion of the construction of the relevant basic Feynman
integrals. For each of the eighteen topologies a {\sc Form} module was created
within which the algorithm to break the original Feynman graphs up into its
basic components was encoded. The tedious identification with the above results
together with the remaining more elementary ones were also contained in each
module. Finally, prior to summing all the results from the $36$ four loop
diagrams, the $\epsilon$ expansion of $I$ and other integrals were evaluated
to the appropriate order in $\epsilon$. The resulting sum produced the 
divergent part of the mass component of the $2$-point function to the simple
pole in $\epsilon$ as a function of the bare parameters.  

\sect{Four loop renormalization.}

The final piece of the calculation rests in determining the overall 
renormalization constant $Z_m$ at four loops in $\MSbar$. However, prior to
this we must consider the full theory. To this point we have tacitly assumed
that only the basic $4$-point vertex of (\ref{baregn}) is responsible for all
the Feynman diagrams we have discussed. The presence of the generated 
evanescent operator in (\ref{rengn}) needs to be included. As noted earlier
since the operator appears with a coupling $g^4$ the effect of this operator
cannot arise before four loops. Therefore, we now have to include the 
additional graph of Figure $7$ where the circle with a cross in it denotes the
insertion of the operator ${\cal O}_3$ with its associated renormalization
constant $Z_{33}$. The integration routine to determine its contribution is the
same as that for the original vertex except that one has to first replace the 
$\Gamma_{(3)}^{\mu\nu\sigma}$ matrices by the corresponding string of 
ordinary $\gamma$-matrices.  

\vspace{0.5cm} 
\begin{figure}[ht]
\hspace{6.5cm} 
\epsfig{file=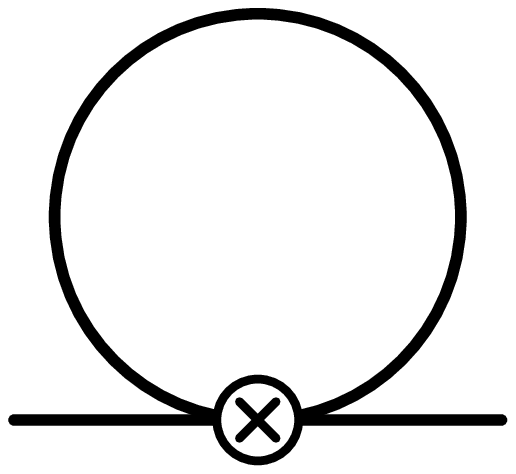,height=2cm} 
\vspace{0.5cm} 
\caption{Contribution from evanescent operator to the four loop mass
renormalization.}
\end{figure} 

With this additional graph included the overall renormalization constant is
extracted using the standard method for automatic Feynman diagram computations
developed in \cite{29}. Briefly one computes the Green's function of interest
as a function of all the bare parameters such as the coupling constant and the 
mass. Then the renormalized parameters are introduced by the rescaling defined
by the renormalization constants. In the present context these are the
renormalization constants leading to the {\em naive} anomalous dimensions as
defined in (\ref{rengn}) and (\ref{rencon}). This rescaling in effect 
reproduces the counterterms to remove subgraph divergences. Moreover the 
Green's function is multiplied by the associated renormalization constant which
in our case and conventions is $Z_\psi Z_m$. As the former is already known,
\cite{9}, then the divergences which remain in the $2$-point function are
absorbed by the unknown pieces of $Z_m$. We recall that at four loops the
anomalous dimension of \cite{9} corresponds to the naive anomalous dimension
$\tilde{\gamma}(g)$ since there is no contribution from the graph of Figure $7$ 
in the wave function channel. Therefore, having followed this procedure we find
the naive mass anomalous dimension in $\MSbar$ is 
\begin{eqnarray}
\tilde{\gamma}_m(g) &=& -~ (2N-1) \frac{g}{2\pi} ~+~ (2N-1) 
\frac{g^2}{8\pi^2} ~+~ (4N-3)(2N-1) \frac{g^3}{32\pi^3} \nonumber \\
&& +~ \left[ ( 48 N^3 - 384 N^2 + 492 N - 138 ) \zeta(3) - 40 N^3 - 72 N^2
+ 160 N - 81 \right] \frac{g^4}{384\pi^4} \nonumber \\
&& +~ O(g^5) ~. 
\label{naigam}
\end{eqnarray}  
At this stage several comments are necessary. First, there are several checks
on the underlying renormalization constant itself. Whilst the evanescent
operator issue arises at four loops, it will manifest itself in the simple pole
in $\epsilon$ of $Z_m$. Therefore, the quartic, triple and double poles in
$\epsilon$ are in fact already predetermined by the structure of previous loop
order poles from the renormalization group equation. For (\ref{naigam}) we have
verified that this is in fact correct.  One other useful check was the explicit
cancellation of divergences of the form $\Delta(0)/\epsilon^n$ for $n$~$=$~$1$
and $2$ at four loops. This is non-trivial since, for instance, $\Delta(0)$ 
arises at three loops both associated with a simple pole in $\epsilon$ and in
the finite part. Therefore, one needs to write $\Delta(0)$ as a formal
expansion in powers of $\epsilon$ prior to the rescaling of the bare 
quantities. This is because the $O(1)$ piece at three loops will be multiplied 
by $1/\epsilon$ poles. Moreover, since $\Delta(0)$ has dependence 
$(m^2)^{d-3}$, then this has to be explicitly factored off since this mass is 
bare and needs to be renormalized too. Once written in this way we have checked 
that the poles in $\epsilon$ involving the $O(1)$ and $O(\epsilon)$ residues 
stemming from the $\epsilon$ expansion of $\Delta(0)$ do indeed cancel 
completely. 

Again one can partially check part of (\ref{naigam}) from another point of 
view. In \cite{9} the structure of the mass anomalous dimension has been given 
in the large $N$ expansion to $O(1/N^2)$ based on the results from a series of 
articles \cite{30,31,32,33,34,35,36}. Again at this level of expansion the 
evanescent operator is not manifested and so the $O(1/N^2)$ coefficients of the
mass anomalous dimension which are given there at four and higher loops in fact
equate to those of the naive mass anomalous dimension $\tilde{\gamma}_m(g)$. In
other words if it were possible to compute the critical exponent corresponding 
to the mass anomalous dimension at the $d$-dimensional fixed point of the 
theory at the next order in large $N$, $O(1/N^3)$, then unless the effect of 
the evanescent operator could be included, it would not correspond to the true 
mass anomalous dimension, \cite{9}. From the expression given in \cite{9} we 
note that when the same convention is used, that part of (\ref{naigam}) at four
loops which corresponds to the $O(1/N^2)$ piece agrees precisely with \cite{9}.
This is a reassuring cross-check on a significant part of our four loop 
computation since, within the computer setup, one can examine the 
$N$-dependence multiplying all the basic integrals which we have had to compute
for all topologies. The vast majority are at least touched by a quadratic or 
cubic in $N$ which are related respectively to the $O(1/N^2)$ or $O(1/N)$ large
$N$ piece already determined in \cite{9}. For the small number of remaining 
pieces which have linear factors in $N$ we have been careful in evaluating the 
corresponding, though invariably simple, vacuum bubbles. Therefore, we are 
confident that (\ref{naigam}) is correct.  
 
One clear problem remains which is related to the structure of the expression
(\ref{naigam}). Unlike the previous orders the four loop part does not vanish
when $N$~$=$~$\half$ which corresponds to the free field theory. Moreover, it
transpires that of the eighteen underlying topologies only the graphs for one
do not vanish for this value for $N$. (Though actually the parts from the 
second and third graphs of Figure $6$ cancel between each other which is
similar to what occurs at three loops for analogous graphs.) The topology which
gives a contribution for $N$~$=$~$\half$ is the final graph of Figure $6$.
However, given our discussion in several places concerning the evanescent 
operator, the resolution is clearly straightforward. More concretely one can 
see the evidence for this if one evaluates (\ref{naigam}) at $N$~$=$~$\half$ to
find  
\begin{equation}
\left. \frac{}{} \tilde{\gamma}_m(g) \right|_{N=\half} ~=~ [ 3 \zeta(3) - 4 ] 
\frac{g^4}{64\pi^4} ~+~ O(g^5) ~. 
\label{gamdisc}
\end{equation}  
This is the piece which needs to be cancelled in order to have a mass dimension
consistent with a free field theory. Indeed this is the relative combination of
rationals and $\zeta(3)$ which our three loop $4$-point function 
renormalization reevaluation produced. Therefore, using (\ref{truerge}) and
(\ref{beta3}) we can derive the true mass anomalous dimension as  
\begin{eqnarray}
\gamma_m(g) &=& -~ (2N-1) \frac{g}{2\pi} ~+~ (2N-1) \frac{g^2}{8\pi^2} ~+~ 
(4N-3)(2N-1) \frac{g^3}{32\pi^3} \nonumber \\
&& +~ \left[ 12 (2N-13)(N-1) \zeta(3) - 20N^2 - 46N + 57 \right] (2N-1)
\frac{g^4}{384\pi^4} \nonumber \\
&& +~ O(g^5) ~. 
\label{truegam} 
\end{eqnarray}  
Clearly this has the correct expected $N$~$=$~$\half$ property and given our
earlier checks on (\ref{naigam}) we will regard (\ref{truegam}) as the 
completion of our original aim. Also, it is worth stressing that the 
discrepancy in the $4$-point function renormalization has now been crucially
resolved simultaneously. It turns out that neither of the previous expressions 
for $\beta_3(g)$, \cite{17,18}, could be correct to preserve the vanishing of 
$\gamma_m(g)$ in the free field case. So we can regard this mass anomalous 
dimension calculation as also a non-trivial check on the full {\em three} loop 
$\MSbar$ renormalization.  

\sect{Discussion.}

We have completed the four loop renormalization of the mass anomalous dimension
of the Gross-Neveu model in the $\MSbar$ scheme. Despite the lack of 
multiplicative renormalizability when the Lagrangian is regularized 
dimensionally, it has been possible to compute an expression which passes all
possible internal checks. Not least of these is the correct implementation of
the projection formula formalism of \cite{13,14} which has been justified by
the consistency with the free field case. Concerning this the previous attempts
to deduce $\beta_3(g)$ appear to indicate that the only approach which is truly
reliable for renormalizing the model is the one where there is a non-zero mass.
This seems to be the conclusion one must draw from the origin of the necessary 
$\zeta(3)$ part missing from \cite{17} required to balance the discrepancy of 
(\ref{gamdisc}). Given these remarks one possible extension would now be to 
repeat the derivation of the mass anomalous dimension at four loops in other 
$4$-fermi models in two dimensions. Whilst considering the most general 
possible interactions involving ${\cal O}_0$, ${\cal O}_1$ and ${\cal O}_2$ of 
\cite{13,14} would perhaps be too ambitious, there is the interesting case of
the non-abelian Thirring model, \cite{37,38}. The seed interaction involves
${\cal O}_1$ but includes colour group generators too. It has been 
renormalized at three loops in $\MSbar$ in \cite{18} and the four loop wave
function is also known, \cite{10}. Though in light of our comments on the
$4$-point function in the Gross-Neveu model, the corresponding $4$-point 
function renormalization would clearly need to be reconsidered to deduce the
correct evanescent operator $\beta$-functions. One motivation for determining 
the mass anomalous dimension in the non-abelian Thirring model would be to 
examine the colour group Casimir structure of the final expression since, given
the similarity with QCD, it is thought that it should involve the same
structures as the corresponding expression for the quark mass anomalous
dimension, \cite{39,40}. This was the case for the wave function, \cite{10}.  

\vspace{1cm}
\noindent
{\bf Acknowledgements.} The author thanks the Max Planck Institute for the
Physics of Complex Systems, Dresden, Germany where part of this work was 
carried out. Also the author is grateful to Dr R. Mertig for assistance with
setting up {\sc Tarcer}.

\end{document}